# Entanglement dynamics driven by topology and non-Hermiticity

Li-Wei Wang[1,2,#], Bolun Hu[3,#,†], Haixiao Zhang[4], Kefan Sun[4], Ying Cheng[4,†], and Jian-Hua Jiang[1,2,5,†]

[1]*State Key Laboratory of Bioinspired Interfacial Materials Science, Suzhou Institute for Advanced Research, University of Science and Technology of China, Suzhou 215123, China*

[2]*School of Physical Sciences, University of Science and Technology of China, Hefei 230026, China*

[3]*School of Science, Jiangsu Provincial Research Center of Light Industrial Optoelectronic Engineering and Technology, Jiangnan University, Wuxi 214122, China*

[4]*Department of Physics, MOE Key Laboratory of Modern Acoustics, Collaborative Innovation Center of Advanced Microstructures, Nanjing University, Nanjing 210093, China*

[5]*School of Physical Science and Technology and Collaborative Innovation Center of Suzhou Nano Science and Technology, Soochow University, Suzhou 215006, China*

[#]These authors contributed equally to this work.

[†]Correspondence should be sent to: jhjiang3@ustc.edu.cn (J.-H.J.), chengying@nju.edu.cn (Y.C.), and bolunhu@jiangnan.edu.cn (B.H.).

## Abstract

**The interplay between topology and non-Hermiticity gives rise to exotic dynamic phenomena that challenge conventional wave-packet propagation and entanglement dynamics. While recent studies have established the non-Hermitian skin effect (NHSE) as a key mechanism for anomalous wave dynamics, a unified framework for characterizing and controlling entanglement evolution in non-Hermitian topological systems remains underdeveloped. Here, by combining theory and experiments, we demonstrate that entanglement entropy (EE) and transport currents serve as robust dynamic probes to distinguish various non-Hermitian topological regimes. Using a generalized non-Hermitian Su-Schrieffer-Heeger model implemented in an acoustic analog platform, we identify three dynamic phases—bulk-like, edge-like, and skin-like—each exhibiting unique EE signatures and transport characteristics. In particular, skin-like dynamics exhibit periodic information shuttling with finite, oscillatory EE, while edge-like dynamics lead to complete EE suppression. We further map the dynamic phase diagram and show that EE scaling and temporal profiles directly reflect the competition between coherent delocalization and NHSE-driven localization. Our results establish a programmable approach to steering entanglement and transport via tailored non-Hermitian couplings, offering a pathway for engineering quantum information dynamics in synthetic phononic, photonic, and quantum simulators.**

# Introduction

Non-Hermitian systems have emerged as a central platform for exploring novel spectral topology and unconventional dynamics across quantum and classical realms[1–7]. Unlike their Hermitian counterparts, non-Hermitian Hamiltonians support complex energy spectra and non-orthogonal eigenstates, giving rise to distinctive phenomena such as exceptional points[2,6,7], nontrivial spectral winding[4,5], and state permutations[8]. In systems with open boundaries, nontrivial spectral winding can lead to non-Hermitian skin effects (NHSEs)[3,9], wherein an extensive number of bulk eigenstates become exponentially localized at the boundaries—breaking the conventional bulk-boundary correspondence and necessitating the description based on generalized Brillouin zones[3,9,10].

An emerging frontier lies in understanding non-Hermitian dynamic processes, where static eigenstate analysis often falls short[11,12]. Complex spectra and non-norm-preserving time evolution lead to rich temporal behaviors, including chiral Zener tunneling[13], anharmonic Rabi oscillations[14], and dynamic NHSEs[15–17]. Such phenomena challenge traditional notions of wave-packet propagation[18,19] and reshape our understanding of quench and transport dynamics beyond unitary frameworks[20–24], thus stimulating much interest in both quantum and classical regimes. Recent experiments on photonic, phononic, and cold-atom platforms have just begun to validate these effects[25–32].

Among the tools characterizing non-Hermitian dynamics, entanglement entropy (EE) has gained considerable attention. In Hermitian systems, EE plays a pivotal role in diagnosing phase transitions, quantum criticality, and many-body thermalization—for instance, revealing the area-law scaling in gapped phases[33] or volume-law (or logarithmic) scaling in gapless phases[34,35] which hold even in free particle systems. In contrast, non-Hermitian settings exhibit dramatically different EE dynamics[36,37]. Recent studies show that NHSEs and non-Bloch dynamics can induce entanglement phase transitions[38–40], where the scaling behavior of EE changes abruptly under parameter tuning[39,41]. These transitions arise from the interplay between non-Hermicity and topological mechanisms. Experimentally, EE has been measured recently in classical analogues such as phononic systems[42], opening the possibility for real-time probe of nonequilibrium topology beyond eigenstates based picture.

Despite these advances, a unified framework for steering and classifying wave dynamics in non-Hermitian systems—particularly using EE as a central probe—remains largely uncharted. A core challenge is to clarify the respective roles of spectral topology, skin-mode localization, and temporal amplification/attenuation as well as their synergy in governing entanglement generation and propagation. Here, we directly tackle this challenge by elucidating how the interplay between topology and non-Hermiticity directs wave-packet dynamics and entanglement flow. We uncover new types of dynamic transitions rooted in this interplay. By doing so, we establish EE as a robust diagnostic for non-Hermitian phases and provide a systematic approach to controlling nonlocal correlations by tailoring non-Hermitian interactions. Furthermore, we identify the mechanisms that govern the flow and redistribution of nonlocal correlations, revealing distinct dynamic regimes with markedly different entanglement behaviors. These findings elucidate the interplay between entanglement, topology, and non-Hermiticity in wave dynamics which can persist to quantum limit for nearly free particles, and is thus inspiring for research on nonequilibrium quantum dynamics in topological systems.

## Results

We start by comparing wavepacket dynamics in one-dimensional (1D) Hermitian and non-Hermitian topological lattices (Figs. 1b and 1c). In Hermitian systems, a localized bulk wavepacket splits into components with different group velocities, which propagate symmetrically, if a sharply local wave packet starts from the center of the system. Upon reaching the boundaries, these components reflect elastically and disperse throughout the lattice according to the bulk dispersion. In contrast, edge modes in Hermitian topological lattices remain tightly localized at the boundaries with minimal spreading into the bulk.

Introducing non-Hermiticity into topological lattices fundamentally alters their dynamic landscape, as schematically contrasted in Figs. 1b and 1c. Under non-Hermitian conditions, wave-packet propagation acquires a pronounced directional bias: bulk components moving rightward are exponentially amplified, whereas the leftward-propagating components attenuate, leading to a net accumulation of energy near the right boundary. Topological edge modes retain their spatial localization but now experience exponential amplification or attenuation dictated by the sign of the imaginary part of their eigenfrequencies. This interplay between non-Hermitian gain/loss processes and topological structure gives rise to a qualitatively distinct paradigm of wave transport and localization, beyond what is known in Hermitian limits.

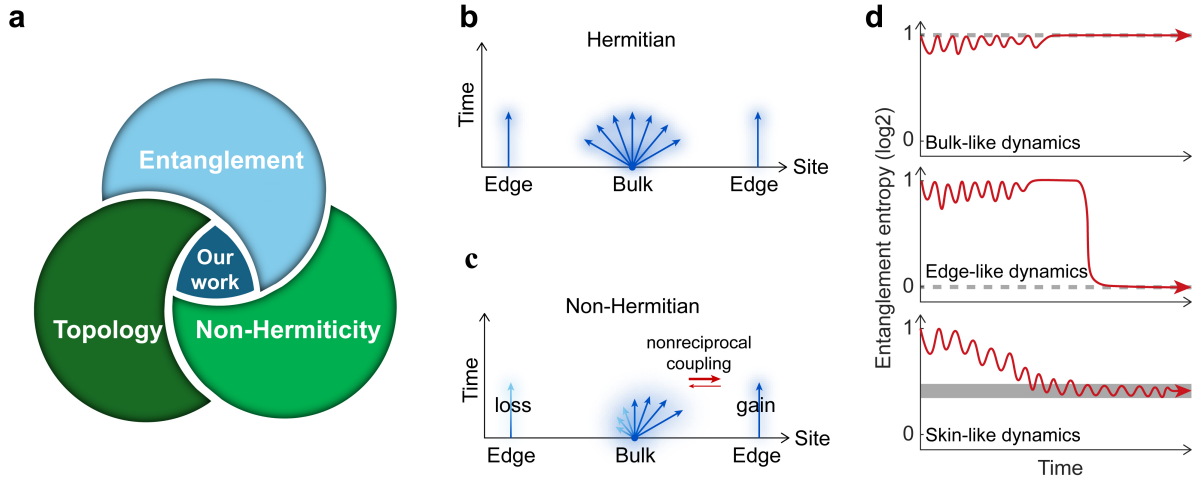

**Fig. 1 | Entanglement dynamics in Hermitian and non-Hermitian topological lattices. a,** Intersection of our work with entanglement, topology, and non-Hermiticity. **b,** Dynamics of Hermitian topological lattices. Wave packets spread symmetrically into the bulk, while topological edge states remain robustly localized at the edges, consistent with energy conservation. **c,** Dynamics of non-Hermitian topological lattices. Gain–loss distribution and nonreciprocal coupling drive asymmetric propagation, leading to directional transport and selective amplification or attenuation of edge and bulk modes. **d,** Characteristics of entanglement entropy evolution in bulk-like, edge-like, and skin-like dynamics. The grey dashed line and the shaded band, respectively, denote the extremal value and the finite interval of the entanglement entropy in the long-time limit.

To systematically characterize how such dynamics influence the nonlocal correlations, we analyze three representative dynamic regimes—bulk-like, edge-like, and skin-like—through the lens of entanglement entropy (EE) evolution. Here, EE is defined via the nonlocal correlation in the left half of the whole system. Thus, it indicates how much mutual information between the left and right halves of the entire system. During the initial propagation stage, both bulk-like and edge-like dynamics exhibit irregular EE oscillations approaching log2, reflecting the persistence of substantial inter-subsystem correlations sustained by Hermitian, diffusion-like, bidirectional wave propagation. In contrast, skin-like dynamics rapidly suppress EE due to strongly nonreciprocal currents that channel waves predominantly into one subsystem. Over longer timescales, these regimes diverge markedly: bulk-like dynamics promote symmetric diffusion in the entire system, stabilizing EE near log2. Edge-like dynamics funnel the wave packet toward a single boundary, thereby reducing EE to zero as all information becomes localized within one subsystem. Skin-like dynamics, dictated by the NHSE, exhibits periodic and stable information shuttling between subsystems, confining EE to a finite interval rather than driving it to an extremal value.

The above behaviors demonstrate that the dynamic architecture of nonlocal correlations is dictated by the underlying transport mechanism—whether Hermitian diffusion, topological boundary trapping, or nonreciprocal skin current—offering a unified framework for understanding entanglement flow in non-Hermitian topological systems.

**Generalized 1D non-Hermitian SSH model**

To elucidate the steering of EE flow due to the interplay between topology and non-Hermiticity, we employ a generalized non-Hermitian Su-Schrieffer-Heeger (SSH) model with open boundary condition (OBC). As illustrated in Fig. 2a, the topological characteristics are governed by intracell and intercell couplings $\kappa_1$ and $\kappa_2$, while non-Hermiticity is introduced through onsite gain/loss rate $\gamma$ and a tunable nonreciprocal coupling $\delta\kappa$. This Hamiltonian is expressed as:

$$\hat{H} = \sum_n \left[ \left( \kappa_1 \hat{a}_n^\dagger \hat{b}_n + \kappa_2 \hat{a}_n^\dagger \hat{b}_{n-1} + h.c. \right) - i\gamma \hat{a}_n^\dagger \hat{a}_n + i\gamma \hat{b}_n^\dagger \hat{b}_n - \delta\kappa \hat{a}_n^\dagger \hat{b}_n \right]. \tag{1}$$

We will show that this model provides a paradigmatic platform to investigate how non-Hermiticity reshapes topological dynamics. The OBC complex energy spectra directly dictate the long time evolution, revealing a rich variety of dynamic behaviors that depend sensitively on both topological character and the degree of non-Hermiticity. In particular, the interplay between gain/loss and non-reciprocal couplings gives rise to distinct dynamic phases that manifest even in topologically trivial regimes, while topologically nontrivial regimes further enrich the dynamic landscape through topological edge trapping and amplification.

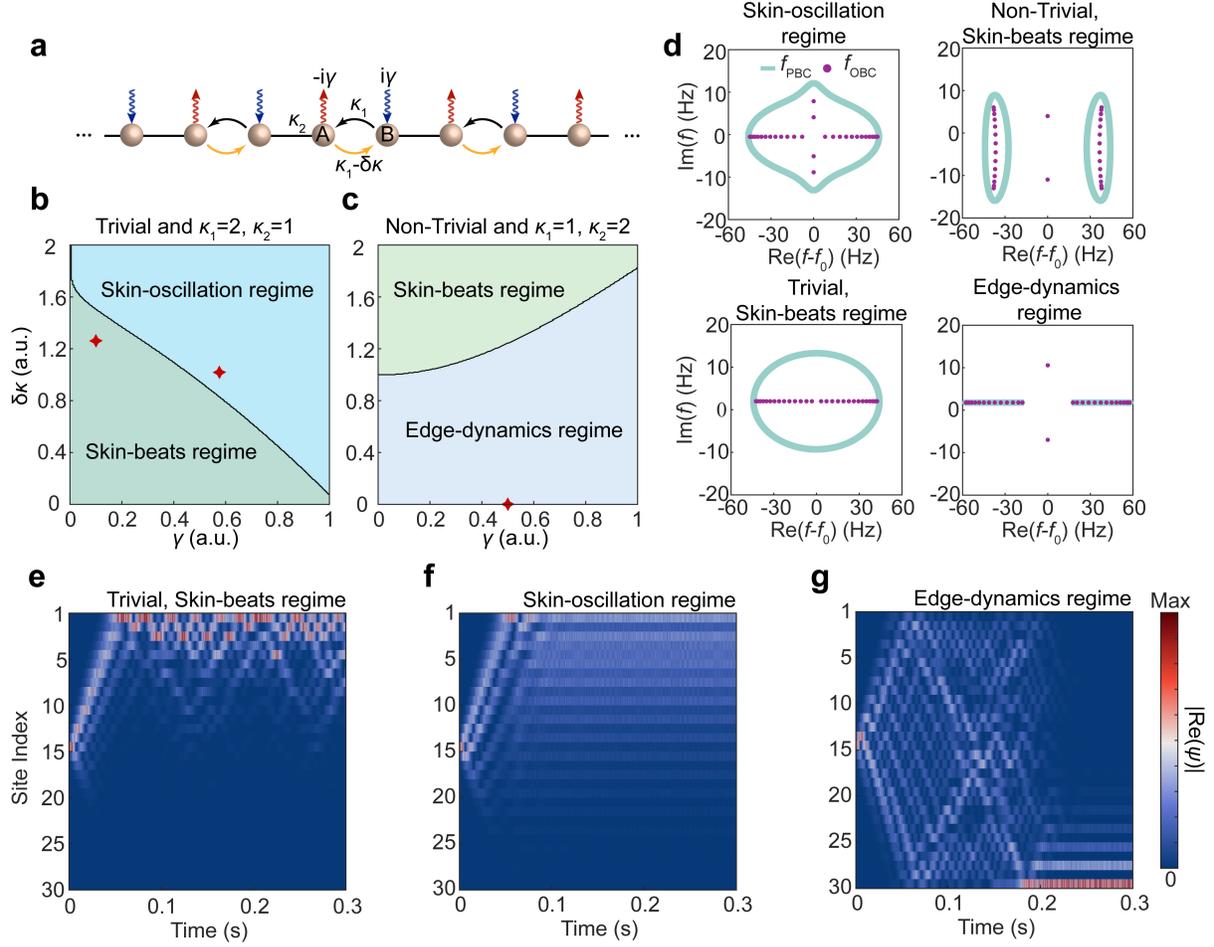

**Fig. 2 | Generalized 1D non-Hermitian SSH model and representative dynamic regimes. a.** Schematic of the generalized 1D non-Hermitian SSH lattice. Light brown spheres denote lattice sites. Reciprocal couplings $\kappa_2$ are indicated by black lines, while non-reciprocal couplings $\kappa_1$ and $\kappa_1 - \delta\kappa$ are represented by black and yellow arrows, respectively. Blue and red wavy arrows correspond to on-site gain and loss. **b, c** Dynamic phase diagrams in the $\delta\kappa$-$\gamma$ plane for the trivial ($\kappa_1 = 2, \kappa_2 = 1$) and non-trivial ($\kappa_1 = 1, \kappa_2 = 2$) cases. Red stars indicate the experimentally equivalent parameters: the trivial cases: ($\delta\kappa = 1.2613, \gamma = 0.1002$) and ($\delta\kappa = 1.0186, \gamma = 0.5759$); the non-trivial case: ($\delta\kappa = 0, \gamma = 0.4406$). In panels **b** and **c**, the parameters $\kappa_1, \kappa_2, \delta\kappa,$ and $\gamma$ are shown in arbitrary units (a.u.). **d,** Corresponding complex energy spectra of four representative cases under PBC and OBC: Skin-oscillation regime ($f_0 = 2130$ Hz, $\kappa_1 = 38.7$ Hz, $\kappa_2 = 19.9$ Hz, $\delta\kappa = 20.27$ Hz, $\gamma_a = -11.96$ Hz, $\gamma_b = 10.96$ Hz), Trivial skin-beats regime ($f_0 = 2130$ Hz, $\kappa_1 = 38.7$ Hz, $\kappa_2 = 19.9$ Hz, $\delta\kappa = 25.1$ Hz, $\gamma_a = -0.997$ Hz, $\gamma_b = 4.985$ Hz), Non-trivial skin-beats regime ($f_0 = 2130$ Hz, $\kappa_1 = 19.9$ Hz, $\kappa_2 = 38.7$ Hz, $\delta\kappa = 24.47$ Hz, $\gamma_a = -10.96$ Hz, $\gamma_b = 3.99$ Hz), Edge-dynamics regime ($f_0 = 2130$ Hz, $\kappa_1 = 19.9$

Hz , $\kappa_2 = 38.7$ Hz , $\delta\kappa = 0$ Hz , $\gamma_a = -6.98$ Hz , $\gamma_b = 10.56$ Hz). **e,f,g,** Spatiotemporal profiles of normalized $|\text{Re}(\psi)|$ from theoretical calculations for representative three cases.

In the trivial regime $|\kappa_2| < |\kappa_1|$, the system exhibits two qualitatively distinct dynamic behaviors. In the Hermitian limit ($\gamma = 0, \delta\kappa = 0$), the OBC spectrum is entirely real, leading to conventional bulk spreading without edge localization. As non-Hermiticity is gradually introduced while maintaining balanced gain and loss, the spectra under different boundary conditions evolve in strikingly different ways (Fig. 2d). Under periodic boundary condition (PBC), the complex spectrum deforms into a loop, marking the emergence of NHSEs. In contrast, the OBC spectrum remains entirely real, though the bulk gap progressively shrinks. The real OBC spectrum ensures the absence of exponential amplification or decay, but the NHSE drives all bulk eigenmodes to localize at the edges. As a result, the long time evolution is dictated by the collective interference of various skin modes, giving rise to multifrequency oscillation characteristics of the skin-beats regime (Fig. 2e). Further increasing $\gamma$ and $\delta\kappa$ drives the OBC spectrum toward complete gap closing, at which point a dynamic phase transition occurs. Beyond this threshold, the non-Hermiticity drives the band-center eigenfrequencies $f_0$ into the complex plane(Fig. 2d), enabling complex spectrum with non-zero imaginary parts. Consequently, the short time dynamics are still governed by the collective contribution of skin modes. At long times, however, the eigenmode with the largest imaginary part is exponentially amplified, and eventually dominates over all other modes, yielding coherent, single mode skin-oscillation regime(Fig. 2f).

In the topological regime $|\kappa_2| > |\kappa_1|$, the OBC spectrum hosts a pair of topological edge states in the gap. In the Hermitian limit ($\gamma = 0, \delta\kappa = 0$), the system exhibits conventional Hermitian dynamics, consistent with the previous analysis. In this Hermitian dyanmics, the edge states play negligible role. Introducing a gain/loss rate $\gamma$ places the system in the unbroken *PT*-symmetric phase, where gain–loss balance is preserved and the bulk spectrum remains entirely real. In contrast, the two edge states acquire opposite imaginary energy. That is, the left edge mode is lossy while the right edge mode is amplifying. In this regime, as shown in Fig. 2g, the short time dynamics is dominated by bulk wave propagation: the initial wave packet splits and propagates bidirectionally, disperse across the lattice in accordance with the underlying band structure—closely resembling Hermitian evolution (see Supplementary Note 2). At longer times, however, the dynamics are controlled by edge states, particularly the right edge state with self-amplification. Incorporating non-reciprocity $\delta\kappa$ gives rise to the NHSEs, leading to complex bulk spectra under OBC. However, as long as the largest imaginary part of the bulk spectrum remains smaller than that of the edge states, no dynamic phase transition occurs and the system stays within the edge-dynamics regime, now accompanied by skin effect features at short times. In contrast, once the maximum imaginary part of the bulk spectrum surpasses that of the edge states, a dynamic phase transition emerges, driving the system into the skin-beats regime. This regime resembles the trivial case but is uniquely characterized by dual frequency beating dynamics because the long time dynamics is dominated by the two bulk modes (in the upper and lower bulk bands, respectively) with the largest imaginery eigenenergy. However, other modes with close imaginery eigenenergy also contribute to the dynamics, leading to complicated fluctuations in the beating pattern of EE time evolution (Detailed dynamic results are provided in the Supplementary Note 2).

**Experimental validation**

We experimentally realize the generalized non-Hermitian SSH model using an acoustic system, with detailed experimental setup provided in Methods. We focus on three representative non-Hermitian dynamic regimes, marked by red stars in Figs. 2b and 2c, and subject them to detailed theoretical and experimental analysis. In all cases, the system is initialized by a broadband pulse at site 15, with sites 1 and 30 serving as the left and right boundaries, respectively. The spatiotemporal acoustic pressure fields, measured for the three representative cases and shown in Fig. 3b, exhibit excellent agreement with the theoretical prediction.

**Entanglement entropy**

We characterize non-Hermitian nonlocal correlations by examining the EE between spatial subsystems, as calculated from the experimental data and presented in Fig. 3e. EE quantifies the competition among the coherent correlation spreading via bulk modes, the topological confining of correlation via edge modes, and the NHSE induced localization of correlation toward the edge boundary. During the initial stage, all three regimes exhibit a rapid transient rise in EE approaching log2, signaling effective correlation spreading as a universal transient behavior. Following the initial transient rise, in both skin -beats and skin-oscillation regimes, the dynamic NHSE drives an efficient unidirectional wave transport that results in a rapid EE drop, reflecting the loss of inter-subsystem correlations due to the NHSE-induced wave localization. As an intriguing example, the edge-dynamics regime demonstrates the transition from coherent delocalized dynamics to edge localized dynamics. In the time-evolution of EE, these features are faithfully manifested in the initial rising, the later stable saturation of EE (accompanied with mesoscopic fluctuations[43]), and the final suppression of EE due to wave trapping into the right edge mode.

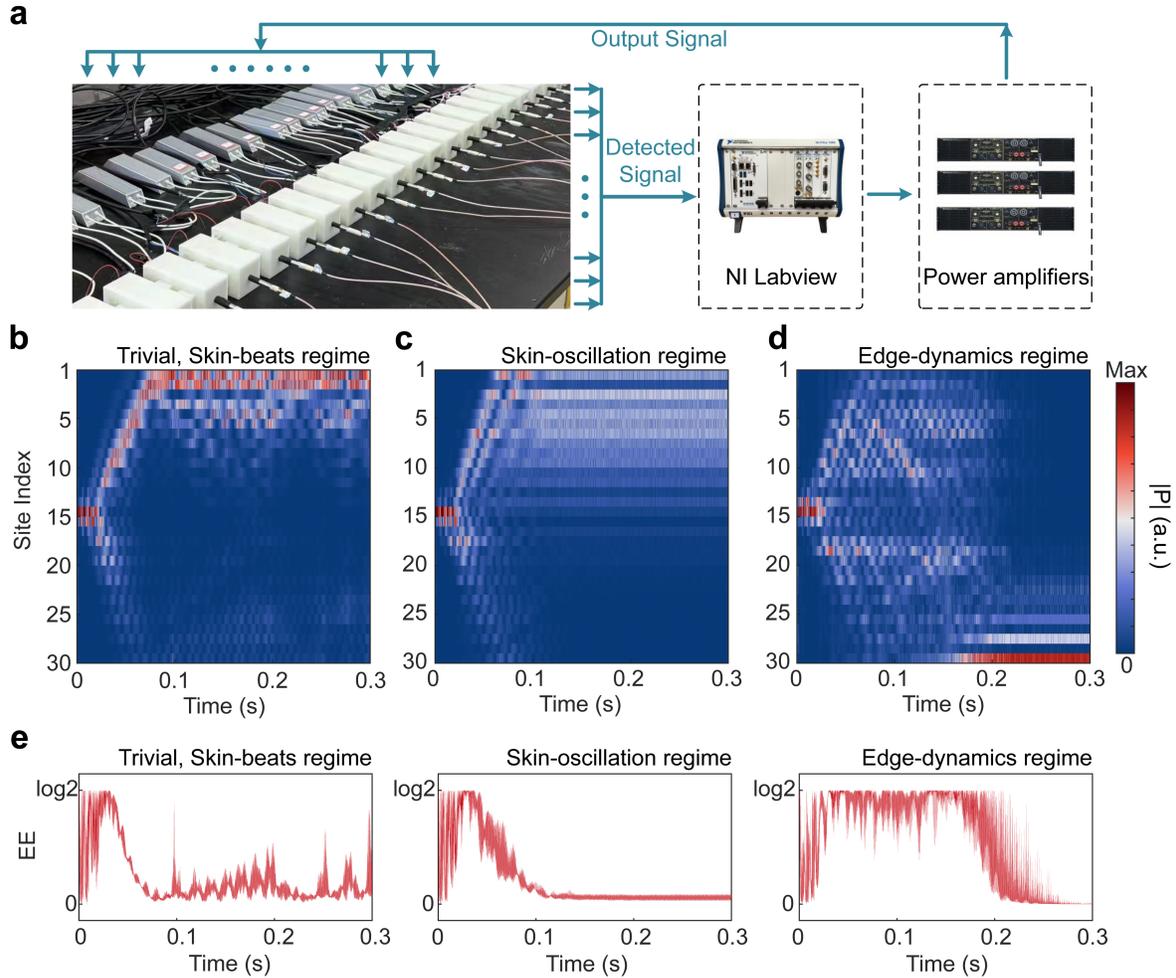

**Fig. 3 | Experimental results, the EE, the local current $J_{Local}$ and total current $I_{Total}$ across three dynamic regimes. a,** Schematic of the generalized non-Hermitian SSH acoustic lattice. **b,c,d** Spatiotemporal pressure profiles $|P|$ for the representative three cases, reconstructed from experimental measurements. The pressure field is obtained from the initial distribution $p(x, t_0)$ and its temporal increments $\Delta p(x, t_i)$ via $p(x, t_j) = p(x, t_0) + \sum_{i=1}^{j} \Delta p(x, t_i) \Delta t$, with $|P|$ normalized at each time step. **e,** The EE extracted from the experimental data for the three cases. Experimental parameters are chosen to match those used in the calculation of Fig. 2e,2f,2g.

**Local currents and total current**

We characterize transport properties by distinguishing between the local current $J_{local}$, which resolves the spatial distribution of probability flow, and the total current $I_{total}$, which quantifies the global transport across the system, as illustrated in Fig.4a. During the initial transport stage, both the skin-beats and skin-oscillation regimes manifest the dynamic NHSEs. In these cases, $J_{local}$ exhibits pronounced unidirectional flow toward the left boundary, leading to a large negative $I_{total}$. In contrast, the edge-dynamics regime is characterized by bidirectional expansion and non-zero $J_{local}$ emerges solely at propagating wavefronts, giving an irregularly oscillating $I_{total}$ driven by bulk modes interference. In the long-time stage, the skin-beats regime displays chaotic oscillation due to multimode interference, yet maintains a negative time averaged total current, indicating persistent wave transfer toward the left boundary. However, the skin-oscillation regime, dominated by a single frequency mode, shows a steadier dynamic NHSE, yielding a larger and temporally stable negative $I_{total}$ without significant fluctuations. Finally, current evolution in the edge-dynamics regime exhibits a unique two-stage process. It starts with a stage dominated by bulk wave propagation and interference, then gradually enters into a stage dominated by the topological edge mode at the right side. Thus, the local currents evolve from a spreading bidirectional pattern to a vanishing pattern as once the wave is trapped at the edge it cannot flow elsewhere. Notably, the crossover between these two stages is driven by the nonreciprocal couplings. Therefore, the time-averaged total current is still positive.

**Entanglement entropy characteristics: non-Hermiticity and system-length scaling**

Here, we characterize nonlocal correlations by examining the dependence of the EE on the non-Hermiticity parameter $\delta\kappa$ across different topological regimes. The EE responds oppositely to increasing $\delta\kappa$ in these regimes. In the trivial regime, the EE is suppressed as increasing $\delta\kappa$ gradually converts extended bulk modes into skin modes, thereby inducing stronger confinement of waves and weaker mutual information. This suppression aligns with the observation of suppressed EE in the long time limit (Fig. 4c). In contrast, the EE exhibits a complex dependence in the topologically nontrivial regime. Specifically, the edge-dynamics regime has near zero EE, indicating robust topological edge trapping in our generalized non-Hermitian SSH model for a range of $\delta\kappa$. Until the largest imaginary parts of the bulk skin modes exceed those of the edges, the skin-beats regime dominates the dynamics, triggering a growth of EE from zero with increasing $\delta\kappa$. Furthermore, the scaling of EE with system size reflects the underlying transport mechanism. The Hermitian case maintains near maximal log2 EE, confirming symmetric, size independent wave dynamics. However, both the skin-beats and skin-oscillation regimes display suppressed EE with increasing size, as the dynamic NHSE causes the wavepacket to become progressively localized near one edge, thereby reducing inter-subsystem correlations. This effect is more pronounced in the skin-beats regime (larger $\delta\kappa$), reflecting tighter localization and lower nonlocal correlations. In contrast, in the edge-dynamics regime, EE gradually increases with system size, because it takes longer time for a longer system to trap all swave components into one edge.

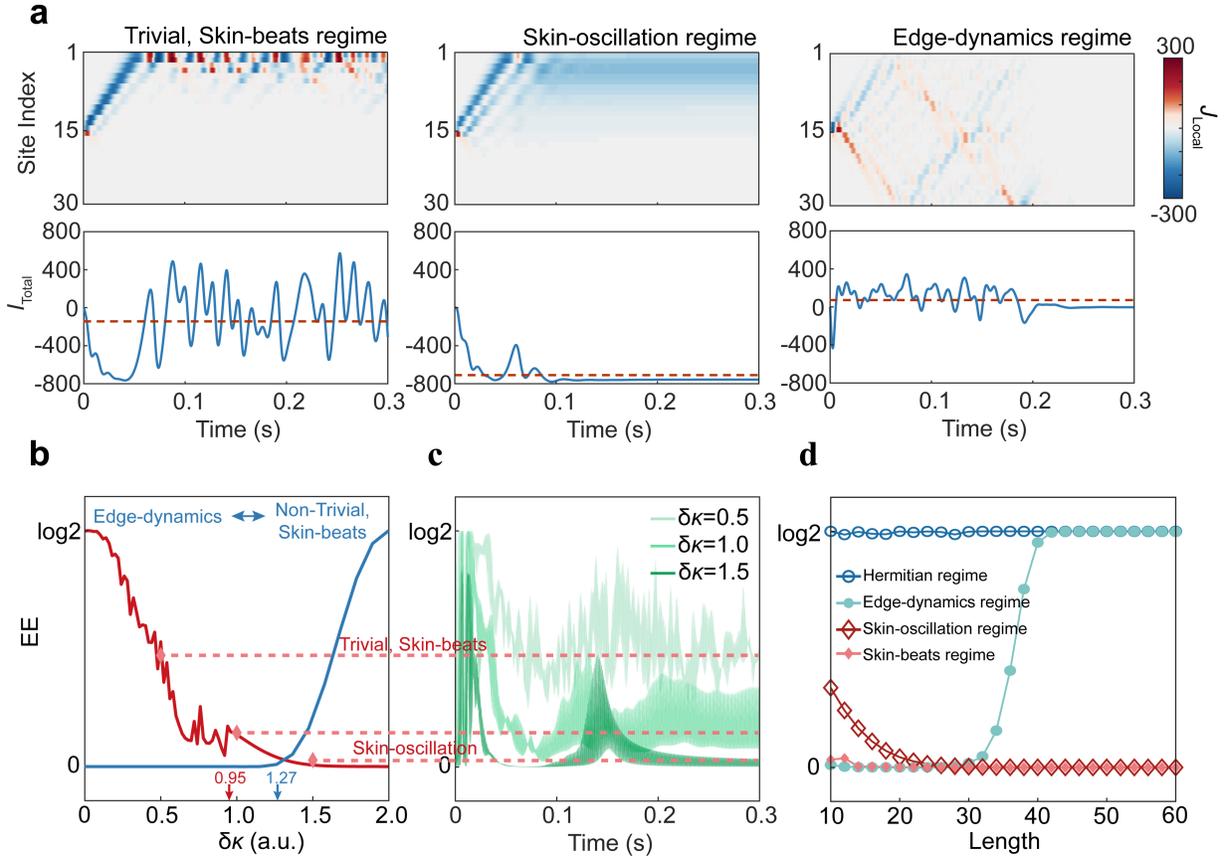

**Fig. 4 | EE for three types of dynamics and its dependence on system length and non-Hermiticity.**
**a,** Corresponding $J_{\text{local}}$ and $I_{\text{total}}$ derived from the calculation for the three distinct dynamic cases. $I_{\text{total}}$ is calculated by summing $J_{\text{local}}$ over all lattice sites, with red dashed lines indicating its time-averaged value. **b,** EE versus non-Hermiticity parameter $\delta\kappa$ at fixed system size $L = 30$ and $t_{a.u.} = 100$s. The dynamics in the trivial regime ($\kappa_1 = 2, \kappa_2 = 1, \gamma = 0.5$) and the non-trivial regime ($\kappa_1 = 1, \kappa_2 =$ $2, \gamma = 0.5$) are shown by the red and blue curves, respectively. Dynamic phase transitions occur at $\delta\kappa = 0.95$ for trivial skin-beats regime and skin-oscillation regime, and $\delta\kappa = 1.27$ for edge-dynamics regime and non-trivial skin-beats regime. Diamond markers indicate three representative experimental parameter sets ($\delta\kappa = 0.5, 1, 1.5$), and the corresponding EE measurements are presented in **c**, with dashed lines highlighting the entanglement values at $t = 0.3s$. **d,** EE is plotted as a function of system length for four representative parameter sets. EE is calculated at $t_{(a.u.)} = 30s$. Parameters are consistent with those presented in Figs. 2b and 2c; the Hermitian case corresponds to ($\delta\kappa = 0, \gamma = 0.4406$). In panels (b) and (d), the parameters $\kappa_1, \kappa_2, \delta\kappa, \gamma$ and $t$ are selected in arbitrary units (a.u.).

## Conclusion and discussion

Our work elucidates how interplay between topology and non-Hermiticity fundamentally steers dynamics and propagation of entanglement in open systems. By mapping the dynamic phase diagram of a generalized non-Hermitian SSH model and experimentally validating it in an acoustic platform, we demonstrate that EE serves as unifying dynamic order parameters for non-Hermitian phases. While spectral topology characterizes the static eigenstates, EE and transport currents capture the coaction between topological and non-Hermitian mechanisms in dynamics, uncovering real-time entanglement flow and current patterns that remain hidden to conventional analysis. Our results also demonstrate that jointly tuning topology and non-Hermiticity enables programmable control over quantum transport and entanglement pathways, offering a novel design principle for engineered functionalities in phononic, photonic, and quantum simulators.

# Methods

## Calculation of dynamic entanglement entropy

We describe a numerical method to investigate the dynamic EE of non-Hermitian system. The dynamic wave function $|\psi(t)\rangle$ evolves from an initial state $|\psi(0)\rangle$ under the Hamiltonian $\hat{H}$ under OBC. The time evolution of the system is governed by the unitary operator $\hat{U}(t) = e^{-i\hat{H}t}$,

which encapsulates all information pertaining to the quantum dynamics and ultimately determines the final state

$$|\psi(t)\rangle = \hat{U}(t)|\psi(0)\rangle. \tag{M1}$$

In particular, the $L \times L$ correlation matrix is

$$C_{ij}(t) = \langle\psi(t)|\hat{c}_i^\dagger \hat{c}_j|\psi(t)\rangle, \tag{M2}$$

where $L$ denotes the freedom degree of system and the dynamic wave function $|\psi(t)\rangle$ is normalized at each time step. Subsequently, we divide the entire system into a subsystem $[x_1, x_2]$ and its complementary part. To calculate the dynamic EE, we diagonalize the $(x_2 - x_1 + 1) \times (x_2 - x_1 + 1)$ submatrix $[C]_{i,j=x_1}^{x_2}(t)$ and obtain eigenvalues $\lambda_n(t)$'s ($n = 1, 2, \ldots, x_2 - x_1 + 1$). Whereupon, the von Neumann EE is given as

$$S(t) = -\sum_{i=1}^{x_2-x_1+1} [\lambda_i \log \lambda_i + (1 - \lambda_i) \log(1 - \lambda_i)]. \tag{M3}$$

In the view of the experiment, spatiotemporal sound pressure $|P|$ can be substituted for dynamic wave function $|\psi(t)\rangle$ in the above equations.

**Calculation of local particle current and total particle current**

We describe the physical derivation for calculating the local particle current $J$ and the total current $I$ in a 1D system. The particle number operators for the left and right systems are denoted as $\hat{N}_n^L$ and $\hat{N}_n^R$, respectively, given by the expressions:

$$\hat{N}_n^{L/R} = \sum_{i \in L/R} \hat{c}_i^\dagger \hat{c}_i. \tag{M4}$$

The local particle current operator of the left and right systems $\hat{J}_n^{L/R}$ satisfies the Heisenberg equation of motion:

$$\hat{J}_n^{L/R} = -\frac{d\hat{N}_n^{L/R}}{dt} = -i[\hat{H}, \hat{N}_n^{L/R}]. \tag{M5}$$

The local particle current $J_n^{L/R}$ is given by the expectation value of the operator of the left and right systems $\hat{J}_n^{L/R}$ in the state $|\psi\rangle$, expressed as

$$J_n^{L/R} \equiv \langle \hat{J}_n^{L/R} \rangle = \langle \psi | \hat{J}_n^{L/R} | \psi \rangle. \tag{M6}$$

The particle current of the single-site system is given by

$$J_n = J_n^L - J_n^R. \tag{M7}$$

Furthermore, the total current $I$ is obtained by summing the particle currents over all particles, given by the expression:

$$I = \sum_n J_n.$$  (M8)

Now, we employ the generalized non-Hermitian SSH model to illustrate the derivation process, where the Hamiltonian of the system can be expressed as

$$H = \sum_n -i\gamma \hat{c}_{n,a}^\dagger \hat{c}_{n,a} + i\gamma \hat{c}_{n,b}^\dagger \hat{c}_{n,b} + \kappa_1 \hat{c}_{n,a}^\dagger \hat{c}_{n,b} + (\kappa_1 - \Delta\kappa)\hat{c}_{n,b}^\dagger \hat{c}_{n,a}$$
$$+ \kappa_2 \hat{c}_{n-1,b}^\dagger \hat{c}_{n,a} + \kappa_2 \hat{c}_{n,a}^\dagger \hat{c}_{n-1,b}.$$  (M9)

The local particle current operator can be calculated as follow:

$$\begin{aligned}
\hat{J}_{a,n}^L &= -\frac{d\hat{N}_{a,n}^L}{dt} = -i\big[\hat{H}, \hat{N}_{a,n}^L\big] \\
&= -i\big[\hat{c}_l^\dagger \hat{c}_k + \hat{c}_k^\dagger \hat{c}_l, \hat{N}_{a,n}^L\big] \\
&= i\kappa_2 \hat{c}_l^\dagger \hat{c}_k - i\kappa_2 \hat{c}_k^\dagger \hat{c}_l
\end{aligned}$$  (M10)

$$\begin{aligned}
\hat{J}_{a,n}^R &= -\frac{d\hat{N}_{a,n}^R}{dt} = -i\big[\hat{H}, \hat{N}_{a,n}^R\big] \\
&= -i\big[\hat{c}_l^\dagger \hat{c}_k + \hat{c}_k^\dagger \hat{c}_l, \hat{N}_{a,n}^R\big] \\
&= i(\kappa_1 - \Delta\kappa)\hat{c}_r^\dagger \hat{c}_k - i\kappa_1 \hat{c}_k^\dagger \hat{c}_r
\end{aligned}$$  (M11)

Then, the particle current of the single-site system $J_n$ can be explicitly computed as

$$\begin{aligned}
J_{a,n}^L \equiv \langle \hat{J}_{a,n}^L \rangle &= \langle \psi(t)|\hat{J}_{a,n}^L|\psi(t)\rangle \\
&= i\kappa_2 \langle \psi(t)|\hat{c}_l^\dagger \hat{c}_k - \hat{c}_k^\dagger \hat{c}_l|\psi(t)\rangle \\
&= i\kappa_2 (\langle \text{vac}|\varphi_l(t)^* \varphi_k(t)|\text{vac}\rangle - \langle \text{vac}|\varphi_k(t)^* \varphi_l(t)|\text{vac}\rangle) \\
&= i\kappa_2 (\varphi_l(t)^* \varphi_k(t) - \varphi_k(t)^* \varphi_l(t)),
\end{aligned}$$  (M12)

where $|\text{vac}\rangle$ is the vacuum state.

$$\begin{aligned}
J_{a,n}^R \equiv \langle \hat{J}_{a,n}^R \rangle &= \langle \psi(t)|\hat{J}_{a,n}^R|\psi(t)\rangle \\
&= \langle \psi(t)|i(\kappa_1 + \Delta\kappa)\hat{c}_r^\dagger \hat{c}_k - i\kappa_1 \hat{c}_k^\dagger \hat{c}_r|\psi(t)\rangle \\
&= i(\kappa_1 - \Delta\kappa)\varphi_r(t)^* \varphi_k(t) - i\kappa_1 \varphi_k(t)^* \varphi_r(t)
\end{aligned}$$  (M13)

$$\begin{aligned}
J_{a,n} \equiv J_{a,n}^L - J_{a,n}^R \\
= i\kappa_2 (\varphi_l(t)^* \varphi_k(t) - \varphi_k(t)^* \varphi_l(t)) - i(\kappa_1 - \Delta\kappa)\varphi_r(t)^* \varphi_k(t) + i\kappa_1 \varphi_k(t)^* \varphi_r(t)
\end{aligned}$$  (M14)

Here, the following equations are used:

$$|\psi(t)\rangle = \sum_i \varphi_i(t)\hat{c}_i^\dagger|\text{vac}\rangle, \qquad \text{(M15)}$$

$$\hat{c}_i|\psi(t)\rangle = \varphi_i(t)|\text{vac}\rangle \qquad \text{(M16)}$$

$$\langle\psi(t)|\hat{c}_i^\dagger = (\hat{c}_i|\psi(t)\rangle)^\dagger = (\varphi_i(t)|\text{vac}\rangle)^\dagger = \langle\text{vac}|\varphi_i(t)^* \qquad \text{(M17)}$$

Therefore, the total current $I_a$ can also be calculated as

$$I_a = \sum_n J_{a,n}. \qquad \text{(M18)}$$

Similarly, by following the same derivation, we obtain the corresponding physical quantities for lattice site $b$, as shown below:

$$J_{b,n}^L \equiv \langle\hat{J}_{b,n}^L\rangle = i\kappa_1\varphi_l(t)^*\varphi_k(t) - i(\kappa_1 - \Delta\kappa)\varphi_k(t)^*\varphi_l(t) \qquad \text{(M19)}$$

$$J_{b,n}^R \equiv \langle\hat{J}_{b,n}^R\rangle = i\kappa_2\big(\varphi_r(t)^*\varphi_k(t) - \varphi_k(t)^*\varphi_r(t)\big) \qquad \text{(M20)}$$

$$\begin{aligned}
J_{b,n} &= J_{b,n}^L - J_{b,n}^R \\
&= i\kappa_1\varphi_l(t)^*\varphi_k(t) - i(\kappa_1 - \Delta\kappa)\varphi_k(t)^*\varphi_l(t) \\
&\quad - i\kappa_2\big(\varphi_r(t)^*\varphi_k(t) - \varphi_k(t)^*\varphi_r(t)\big)
\end{aligned} \qquad \text{(M21)}$$

$$I_b = \sum_n J_{b,n}. \qquad \text{(M22)}$$

Summarizing the results, the local current $J_n$ and total current $I$ is given by

$$J_n = J_{a,n} + J_{b,n}, \qquad \text{(M23)}$$

$$I = I_a + I_b = \sum_n J_n. \qquad \text{(M24)}$$

**Experimental Methods**

The schematic shown in SFig.4 illustrates the experimental implementation of the non-reciprocal intra-cell coupling and onsite gain/loss. We employ micro loudspeakers embedded in the cavity cover plate as acoustic sources. The non-Hermitian factors in the system are controlled by a programmed NI LabVIEW system and power amplifiers. The loudspeaker generate an air volume flow, with its intensity is determined by the sound pressure at corresponding front-end.

Specifically, when introducing onsite gain/loss, the air volume flow follows: $Q_1 = k_0 \cdot \alpha_1 \cdot p_a$. When introducing intra-cell nonreciprocal coupling, the air volume flow follows: $Q_2 = ik_0 \cdot \alpha_2 \cdot p_b$. Here, $k_0$ is determined by the acoustic-electric/electro-acoustic transduction coefficients of the microphones/loudspeakers and the amplification factor of the power amplifier. And $\alpha_{1,2}$ simulate the onsite modulation strength and the nonreciprocal coupling strength.

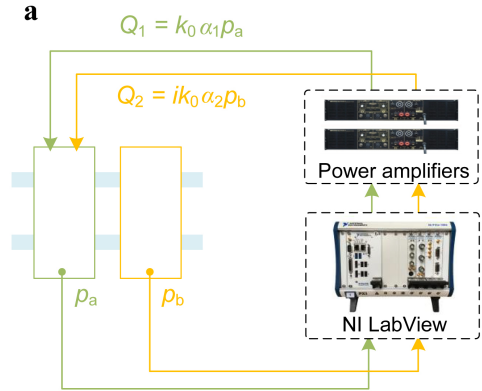

**Method Fig. 1 | Experimental implementation of the acoustic non-Hermiticity.**

During the experiment, we constructed both trivial and nontrivial sonic crystals with highprecision 3D printed structure and integrated non-Hermitian components. The excitation signals were controlled by two clock-synchronized arbitrary waveform generators (NI PXIe-6739) through LabVIEW program, and amplified by the power amplifiers (YAMAHA P-5000). The electrical signals were calibrated to account for the phase and amplitude offset caused by the asynchronous power amplifiers. The initial excitation source was set in No. 15 cavity, with the frequency rage of 2030 Hz- 2230 Hz. The local pressures and phases were measured using 1/4-inch condenser microphones (GRAS 40PH) which were placed at the designated holes. The output signals from the microphones were acquired using two digitizers (NI PXI-4499).

## Acknowledgments

This work was supported by the National Key R&D Program of China (2022YFA1404400), the National Natural Science Foundation of China (12125504 and 12225408), the "Hundred Talents

Program" of Chinese Academy of Sciences, and the Priority Academic Program Development (PAPD) of Jiangsu Higher Education Institutions.

## Author contributions

J.-H.J. conceived the idea. J.-H.J. and Y.C. guided the research. L.-W.W. and J.H.J. proposed and established the theory. B.H., H.Z., K.S. and Y.C. designed and performed the experiments. L.-W.W., B.H., J.-H.J. and Y.C. wrote the main text. B.H. and Y.C. wrote Supplementary Information. All authors were involved in the discussions of the results.

## Competing Interests

The authors declare that they have no competing financial interests.

## Data availability

All data are available in the manuscript and Supplementary Information. Additional information is available from the corresponding authors through request.

## Code availability

Requests for the computation details can be addressed to the corresponding authors.